\documentclass[sn-mathphys,Numbered]{sn-jnl}

\usepackage{graphicx}
\usepackage{float}
\usepackage{multirow}%
\usepackage{amsmath,amssymb,amsfonts}
\usepackage{amsthm}
\usepackage{mathrsfs}
\usepackage[title]{appendix}
\usepackage{xcolor}
\usepackage{textcomp}
\usepackage{manyfoot}%
\usepackage{booktabs}%
\usepackage{algorithm}
\usepackage{mathtools}
\usepackage{algorithmicx}%
\usepackage{algpseudocode}%
\usepackage{listings}%
\usepackage{lmodern}
\usepackage[capitalize]{cleveref}
\usepackage{anyfontsize}
\usepackage{hyperref}
\usepackage{cleveref}

\begin{document}

\title{The Spectroscopy of Kerr-Einstein-Maxwell-Dilaton-Axion: Exact Quasibound States, Scalar Cloud, Horizon's Boson Statistics and Superradiance}

\author[1]{David Senjaya} \email{davidsenjaya@protonmail.com} 
\affil[1]{High Energy Physics Theory Group, Department of Physics, Faculty of Science, Chulalongkorn University, Bangkok 10330, Thailand}

\author[2]{Supakchai Ponglertsakul}
\email{supakchai.p@gmail.com}
\affil[2]{Strong Gravity Group, Department of Physics, Faculty of Science, Silpakorn University, Nakhon Pathom 73000, Thailand}


\maketitle
\begin{abstract}

In the present study, we investigate the quasibound states, scalar cloud and superradiance of relativistic scalar fields bound to a rotating black hole in Einstein-Maxwell-Dilaton-Axion theory (Kerr-EMDA). We present the exact eigensolutions of the governing Klein-Gordon equation in the black hole background. By imposing boundary conditions on the quasibound states, we are able to find the exact complex quasibound state frequencies of the corresponding radial wave functions in terms of the confluent Heun polynomial. Considering light scalar field limit of the obtained solution, we investigate the scalar-black hole resonance configuration known as the scalar cloud. In addition, we obtain analytic relation between light scalar mass and black hole spin for scalar cloud. We explore a boson distribution function by linearly expanding the radial wave function near the black hole's event horizon. Moreover, by applying the Damour-Ruffini method, this allows us to calculate the Hawking radiation flux. In the final section, we consider propagating wave in a slowly rotating Kerr-EMDA black hole for bosons having much larger Compton wavelength comparing to the size of rotating black hole. This condition allows us to use the asymptotic matching technique to calculate the amplification factor for scalar fields in the Kerr-EMDA black hole. We present the dependence of amplification factor on black hole parameters by graphical analysis.

\end{abstract}

\section{Introduction}
An analytical solution to a wave equation in a curved spacetime is extremely significant. There are numerous applications in black hole physics gain a benefit of having an exact (or at least analytical) solution to the wave equation. For example, the study of the greybody factor, scalar cloud, quasinormal modes, entropy and area quantization, quasibound states and superradiance \cite{Lei,Siqueira,Noda,senjaya1,senjaya2}. In general, scalar perturbations are characterized by a discrete spectrum of complex frequencies known as the quasiresonance frequency, with the real component determining the oscillation timescale and the imaginary part determining the exponential decay (stability) or growth (instability) timescale. Consequently, quasiresonance frequency plays an important role in the research of black hole stability under particular perturbation.

The quasibound states are quantized relativistic bound states that are located in the black hole's finite gravitational potential well and outside of its event horizon. As a result, the quasibound states are leaking and moving into the black hole, giving rise to complex-valued frequencies in the spectrum. The imaginary part of the spectrum determines the system's stability, while the real part is linked to the scalar's energy \cite{Vieira:2021doo,Senjaya:2024blm}. Various quasibound states in black hole and analog black hole spacetimes have been reported in the following works: the static dilatonic black hole \cite{Yang}, static black hole's in Einstein–Gauss–Bonnet Gravity \cite{100}, Kerr–Newman black hole \cite{vier21}, Lense-Thirring black hole \cite{Senjaya3}, static black holes of $f(R)$ gravity \cite{senjaya4}, Ernst black hole \cite{senjaya5}, static black hole of Einstein–Maxwell-dilaton (EMD) theory \cite{Senjaya:2024uqg},  optical black hole analog \cite{Senjaya:2024xqm}, The Kerr–Bumblebee black hole \cite{Senjaya:2024blm}, Kerr-Einstein–Yang–Mills–Higgs’s black hole \cite{Senjaya:2024gfh}, dyonic Kerr–Sen black hole \cite{Senjaya:2024gpb} and analog Schwarzschild black holes of Bose-Einstein condensates \cite{Vieira:2023ylz}.

It is found in \cite{Hod:2012px} that there exists infinitely long-lived scalar configuration around maximally rotating black hole. These states, which decay exponentially at asymptotic infinity, have zero imaginary part of the frequency and real part matches exactly at the threshold of superradiance frequency. Neutral and charged scalar clouds on Kerr and Kerr-Newman spacetime are studied in \cite{Benone:2014ssa}. The scalar quasibound states and the scalar clouds configurations require the scalar relativistic energy to be less than the scalar's rest energy \cite{Senjaya:2024rse,Hod:2024aen}. Following the investigation of the black hole's scalar bound states, the Damour-Ruffini approach \cite{Damour,Senjaya:2024uqg,senjaya5} can be used to investigate the Hawking radiation spectra and radiation flux released by the black hole's event horizon.

In contrast to black hole quasibound states and scalar clouds, superradiance is a resonant scattering phenomenon, where the scalar relativistic energy is greater than the scalar rest energy. In a gravitational system, the scattering of radiation off absorbing rotating objects produces waves with amplitudes greater than the incident one under certain conditions, known as rotational superradiance \cite{Franzin,Khodadi:2020cht}. Recent works on superradiance scattering by black holes have been reported in \cite{s1,Lei,Siqueira,Huang:2016qnk,Wu:2024fvy,Jha:2022tdl,Yang:2022uze,Alexander:2022avt,Richarte:2021fbi}. 

In this work, we are interested to study the scalar quasibound states, scalar clouds, Hawking radiation and superradiance scattering of the charged rotating Kerr-EMDA black hole \cite{Garcia}. This paper is organized as follows. In Sect.~\ref{sect:BH}, we introduce stationary axially symmetric black hole solution of the EMDA theory. In Sect.~\ref{sect:KG}, we present the derivation of exact solutions to the Klein-Gordon in the Kerr-EMDA black hole spacetime and the quasibound states are discussed. We consider scalar cloud configuration in Sect.~\ref{sect:cloud}. We discuss boson statistics near the event horizon and determine the energy flux of Hawking radiation in terms of the Jonqui\'ere's Polylogarithm function in Sect.~\ref{sect:Hawking}. In Sect.~\ref{sect:suprad}, we discuss superradiant scattering of scalar fields by the Kerr-EMDA black hole. The superradiance amplification factor is calculated analytically via the matching technique. We conclude our work in Sect.~\ref{sect:conclud}.


\section{Rotating Black Hole in The EMDA Theory}\label{sect:BH}

The low-energy effective theory for heterotic string theory involves reducing and truncating the string theory in four dimensions by the compactification of six of the ten dimensions of the string theory and the omission of the arising massless fields in the obtained heterotic structure. In the compactified theory, only $U(1)$ charges are permitted. The resultant theory consists of the Einstein-Hilbert action, dilaton kinetics term, Maxwell-dilaton coupling, Maxwell-axion coupling and dilaton-axion coupling. The action is defined as follows \cite{Garcia},
\begin{align}
S_{EMDA} &=\int\left(R+\mathcal{L}_m\right)\sqrt{-g}d^4x, \label{action}
\end{align} 
where $R$ is the Ricci scalar and $g$ is the determinant of the metric tensor. The matter Lagrangian is,
\begin{align}
    \mathcal{L}_m &= -2\partial_\mu\varphi\partial^\mu\varphi-\frac{1}{2}e^{4\varphi}\partial_\mu\kappa\partial^\mu\kappa -e^{-2\varphi} F_{\mu\nu}F^{\mu\nu}-\kappa F_{\mu\nu}\tilde{F}^{\mu\nu}.
\end{align}
The Maxwell electromagnetic tensor is defined as $F_{\mu\nu}=\partial_\mu A_\nu-\partial_\nu A_\mu$. The Maxwell electromagnetic tensor's dual is $\tilde{F}^{\mu\nu}=-\frac{1}{2}\sqrt{-g}\varepsilon_{\mu\nu\alpha\beta} F^{\mu\nu}$. There are two dynamical scalar fields: dilaton $\varphi$ and axion $\kappa$.





The theory \eqref{action} admits stationary axially symmetric black hole solution. In Boyer-Lindquist coordinate, 
the line element of the Kerr-EMDA spacetime is given by \cite{Garcia},
\begin{align}
ds^2 &=-\left(1-\frac{rr_s}{\rho^2}\right)dt^2-\frac{2rr_sa{\sin ^2 \theta\ }}{\rho^2}dtd\phi + \frac{\rho^2}{\Delta }dr^2+\rho^2d\theta^2+A\left(r,\theta \right)\frac{{{\sin }^2 \theta\ }}{\rho^2}d\phi^2, \label{metric}  
\end{align}
with,
\begin{gather}
\rho^2=r\left(r-2D\right)+a^2{{\cos }^{{\rm 2}} \theta\ },\\
\Delta =r\left(r-2D\right)-r_sr+a^2,\\
A\left(r,\theta \right)=\left(r(r-2D)+a^2\right)^2-\Delta a^2 \sin^2{\theta}.   
\end{gather}
The $r_s$ denotes Schwarzschild's event horizon $r_s=2M$. The metric above is characterized by three parameters i.e., mass $M$, angular momentum per unit mass $a$ and dilaton parameter $D$. It can be shown that dilaton parameter $D$ and the axion field depend on electric charge of black hole \cite{Ganguly:2014pwa,Sahoo:2023czj}. In addition, the presence of the axion field yields non-zero angular momentum of black hole \cite{Ganguly:2014pwa,Sahoo:2023czj}. Clearly, with zero dilatonic parameter $D=0$, \eqref{metric} reduces to the Kerr solution.

The black hole horizons are found by solving the  $\Delta =0$. The solutions of the quadratic equation represent the outer and inner horizon respectively $r_+,r_-$ as follows,
\begin{gather}
r_{\pm }=\frac{r_s}{2}+D\pm \sqrt{\left(\frac{r_s}{2}+D\right)^2-a^2}. \label{rpm}
\end{gather}

\section{The Klein-Gordon Equation}\label{sect:KG}
A relativistic scalar field in a curved space-time, regardless massive or massless, is represented by the covariant Klein-Gordon equation. The relativistic massive scalar wave equation reads as follows,
\begin{gather}
\left[\frac{1}{\sqrt{-g}}{\partial }_\mu\left(\sqrt{-g}g^{\mu \nu}{\partial }_\nu\right)-{\mu^2}\right]\psi=0,
\end{gather}
where $\mu$ is the scalar field mass and $g$ is determinant of metric tensor i.e., $det(g_{\mu\nu})=g$.

Due to the nature of the temporal and azimuthal symmetry, we can use the following separation ansatz,
\begin{gather}
\psi\left(t,r,\theta,\phi\right)=e^{-iE t+im_\ell\phi}R\left(r\right)T\left(\theta\right),
\end{gather}
where $E$ is the scalar field's energy.
Let us define two dimensionless energy parameters,
\begin{gather}
    \omega=E r_s, ~~~~
    \omega_0=\mu r_s.
\end{gather}
Therefore, we obtain the full radial-angular equation as follows,
\begin{multline}
\left[\frac{1}{{T \sin  \theta\ }}{\partial }_\theta\left({\sin  \theta\ }{\partial }_\theta T\right)-\frac{m^2_\ell }{{{\sin }^2 \theta\ }}-\left(\frac{\omega^2_0a^2}{r^2_s}-\frac{\omega^2a^2}{r^2_s}\right){{\cos }^2 \theta\ }\right]\\
+\left[\frac{1}{R}{\partial }_r\left(\Delta {\partial }_rR\right)+\frac{\omega^2}{r^2_s}\frac{{\left(r\left(r-2D\right)+a^2\right)}^2}{\Delta }-\frac{\omega^2a^2}{r^2_s}+\frac{m^2_\ell  a^2}{\Delta }\right. \\ \left.-2\frac{\left(r\left(r-2D\right)+a^2-\Delta\right)a}{\Delta }\left(\frac{\omega m_\ell}{r_s}\right)-\frac{\omega^2_0}{r^2_s}r\left(r-2D\right)\right]=0.
\end{multline}
The angular part can be treated separately,
\begin{equation}
  \frac{1}{{T\sin  \theta\ }}{\partial }_\theta\left({\sin  \theta\ }{\partial }_\theta T\right)-\frac{m^2_\ell }{{{\sin }^2 \theta\ }}
-\left(\frac{\omega^2_0a^2}{r^2_s}-\frac{\omega^2a^2}{r^2_s}\right){{\cos }^2 \theta\ }+\lambda^{m_\ell}_\ell =0,   \end{equation}
where $\lambda^{m_\ell}_\ell $ is the separation constant. In non-rotating case, it is clear that  $\lambda^{m_\ell}_\ell=\ell\left(\ell+1\right)$ and the angular wave solution is the Legendre polynomial, $P^{m_\ell}_\ell  (\cos \theta)$. In a more general case with $a\neq 0$, the angular solution is represented by the spheroidal harmonics, $S^{m_\ell}_\ell$, 
\begin{gather}
T(\theta){\rm =}S^{m_\ell}_\ell \left(\sigma,{\cos  \theta\ }\right)=\sum^{\infty }_{r=-\infty }{d^{\ell m_\ell}_r\left(\sigma\right)P^{m_\ell}_{\ell+r}\left({\cos  \theta\ }\right)},
\end{gather}
where,
\begin{equation}
 \sigma=   \frac{\omega^2_0a^2}{r^2_s}-\frac{\omega^2a^2}{r^2_s},
\end{equation}
and the coefficient $d^{\ell m_\ell}_r$ is the required weight of the mode $\left(\ell+r\right), m_\ell$ to construct the Spheroidal Harmonics $S^{m_\ell}_\ell$. For a relatively small $\sigma$, the separation constant $\lambda^{m_\ell}_\ell$ can be calculated perturbatively \cite{Press,Berti1,Berti2,Cho:2009wf,Suzuki:1998vy,Ponglertsakul:2020ufm,Breuer} and can be expressed as, 
\begin{equation}
\lambda^{m_\ell}_\ell=\ell(\ell+1)-2\sigma\left(\frac{m_\ell^2+\ell(\ell+1)-1}{(2\ell-1)(2\ell+3)}\right)+O\left(\sigma^2  \right).
\end{equation}
For the sake of notation simplicity, let us define the eigensolution of the angular wave as follows,
\begin{equation}
    S_\ell^{m_\ell}(\theta,\phi)\equiv e^{im_\ell\phi}T(\theta),
\end{equation}
and consequently, the complete wave function is given by,
\begin{gather}
\psi\left(t,r,\theta,\phi\right)=e^{-i\frac{\omega}{r_s} t}R\left(r\right)S_\ell ^{m_\ell}(\theta,\phi).
\end{gather}

\subsection{The Radial Equation}
Here, we continue considering the radial equation. Let's introduce the following short-handed notation,
\begin{align}
{\Delta} &\equiv (r-r_-)(r-r_+),\\
\frac{r_+-r_-}{{\Delta}} &\equiv \frac{\delta_r}{{\Delta}}=\frac{1}{r-r_+}-\frac{1}{r-r_-},  \label{radialeq1} \\
K^{m_\ell}_{\ell} &\equiv \frac{\omega^2}{r^2_s}a^2-\frac{\omega^2_0}{r^2_s}a^2-2\frac{\omega m_\ell a}{r_s}+\lambda^{m_\ell}_{\ell} \label{K}.
\end{align}
We can now rearrange the radial equation as follows,
\begin{multline}
\delta^2_r{\partial }^{{\rm 2}}_rR{\rm +}\left[\frac{1}{r-r_+}+\frac{1}{r-r_-}\right]\delta^2_r{\partial }_rR\\{\rm +}\left[{\left(\frac{1}{r-r_+}-\frac{1}{r-r_-}\right)}^2{\left\{\frac{\omega}{r_s}\left(r\left(r-2D\right)+a^2\right)-{m_\ell a}\right\}}^2\right. \\ \left.-\delta_r\left(\frac{1}{r-r_+}-\frac{1}{r-r_-}\right)\left\{\frac{\omega^2_0}{r^2_s}\left(r\left(r-2D\right)+a^2\right)+K^{m_\ell}_\ell \right\}\right]R{\rm =0}. \label{radialmod}
\end{multline}
As the region of interest is outside the outer horizon, i.e., $r_+\leq r <\infty$, let us use the following new variables,
\begin{gather}
x \equiv r-r_+ \equiv \delta_ry ~~~\to~~~ dx=dr=\delta_rdy.
\end{gather}
This shifts the region of interest from $r$ to $0 \leq y < \infty$. Moreover, we can express $r-r_-$ as $\delta_r(y+1)$. In the terms of $y$, the radial equation \eqref{radialmod} becomes,
\begin{equation}
  {\partial }^{{\rm 2}}_yR{\rm +}\left[\frac{1}{y}+\frac{1}{y+1}\right]{\partial }_yR{\rm +}\left[\mathcal{F}  + \mathcal{G}\right]R{\rm =0},  \label{eq:radialinY}
\end{equation}
where,
\begin{align}
\mathcal{F} &= \left[\frac{1}{\delta_r}\left(\frac{1}{y}-\frac{1}{y+1}\right) \left\{\frac{\omega}{r_s}\left({\left(\delta_ry+r_+\right)}^2-2D\left(\delta_ry+r_+\right)+a^2\right)-{m_\ell a}\right\}\right]^2,  \\  
\mathcal{G} &= -\left(\frac{1}{y}-\frac{1}{y+1}\right)\left\{\frac{\omega^2_0}{r^2_s}\left({\left(\delta_ry+r_+\right)}^2-2D\left(\delta_ry+r_+\right)+a^2\right)+K^{m_\ell}_\ell \right\}.    
\end{align}
Now let us define the following constants,
\begin{align}
    K_1 &\equiv \frac{\omega}{r_s}\left(r_+\left(r_+-2D\right)+a^2\right)-{m_\ell a}, \\
    K_2 &\equiv \frac{\omega^2_0}{r^2_s}\left(r_+\left(r_+-2D\right)+a^2\right)+K^{m_\ell}_\ell , \\
    K_3 &\equiv \left(r_++r_--2D\right)\frac{\omega}{r_s}-\frac{K_1}{\delta_r}, \\
    K_4 &\equiv \frac{\omega^2_0}{r^2_s}\delta^2_r-2\delta_r\left(r_+-D\right)\frac{\omega^2_0}{r^2_s}+K_2.
\end{align}
Therefore the variable $\mathcal{F}$ and $\mathcal{G}$ can be written in terms of above parameters as,
\begin{align}
\mathcal{F} &= \frac{\omega^2}{r^2_s}\delta^2_r+\frac{K^2_1}{\delta^2_ry^2}+\frac{K^2_3}{{\left(y+1\right)}^2}+\frac{2\omega K_1}{r_sy}
+\frac{2\omega\delta_rK_3}{r_s\left(y+1\right)}+\frac{2K_1K_3}{\delta_r}\left(\frac{1}{y}-\frac{1}{y+1}\right), \\
\mathcal{G} &= -\frac{\omega^2_0}{r^2_s}\delta^2_r+\frac{K_4}{y+1}-\frac{K_2}{y}.   
\end{align}
Now, we are ready to convert the radial wave equation into its normal form (see Appendix \ref{AppendixA} for details). Recall that, the radial wave equation \eqref{eq:radialinY} is in the form,
\begin{multline}
{\partial }^{{\rm 2}}_yR{\rm +}\left[\frac{1}{y}+\frac{1}{y+1}\right]{\partial }_yR{\rm +}\left[\frac{\omega^2}{r^2_s}\delta^2_r+\frac{K^2_1}{\delta^2_ry^2}+\frac{K^2_3}{{\left(y+1\right)}^2}+\frac{2\omega K_1}{r_sy}+\frac{2\omega\delta_rK_3}{r_s\left(y+1\right)}\right. \\ \left.+\frac{2K_1K_3}{\delta_r}\left(\frac{1}{y}-\frac{1}{y+1}\right)-\frac{\omega^2_0}{r^2_s}\delta^2_r+\frac{K_4}{y+1}-\frac{K_2}{y}\right]R{\rm =0}.
\end{multline}  
This equation can be recast into its normal form by introducing,
\begin{equation}
Y(y)=\sqrt{y(y+1)}R(y).    
\end{equation}
Thus we obtain,
\begin{equation}
\frac{d^2Y(y)}{dy^2}+K(y)Y(y)=0,
\end{equation}
where the following are identified,
\begin{align}
K(y) &=-\frac{1}{2}\frac{dp}{dy}-\frac{1}{4}p^2+q, \\
p &= \frac{1}{y}+\frac{1}{y+1}, \\
q &=\frac{\omega^2}{r^2_s}\delta^2_r+\frac{K^2_1}{\delta^2_ry^2}+\frac{K^2_3}{{\left(y+1\right)}^2}+\frac{2\omega K_1}{r_sy}+\frac{2\omega\delta_rK_3}{r_s\left(y+1\right)}+\frac{2K_1K_3}{\delta_r}\left(\frac{1}{y}-\frac{1}{y+1}\right) \nonumber \\
&~~~~-\frac{\omega^2_0}{r^2_s}\delta^2_r+\frac{K_4}{y+1}-\frac{K_2}{y},
\end{align}
or explicitly, 
\begin{align}
K(y)&=-\left(\frac{{\omega}^2_0}{r^2_s}\delta^2_r-\frac{{\omega}^2}{r^2_s}\delta^2_r\right)+\frac{1}{y}\left(-\frac{1}{2}+\frac{2{\omega}K_1}{r_s}+\frac{2K_1K_3}{\delta_r}-K_2\right)\nonumber \\ &\phantom{=}+\frac{1}{y^2}\left(\frac{1}{4}+\frac{K^2_1}{\delta^2_r}\right)+\frac{1}{y+1}\left(\frac{1}{2}+\frac{2{\omega}\delta_rK_3}{r_s}-\frac{2K_1K_3}{\delta_r}+K_4 \right)\nonumber \\ &\phantom{=}+\frac{1}{{\left(y+1\right)}^2}\left(\frac{1}{4}+K^2_3\right).
\end{align}
Comparing with the $K(x)$ of confluent Heun's differential equation (see Appendix \ref{AppendixA}), the confluent Heun's parameters can be solved as follows,
\begin{align}
y &= -x,\\
\alpha_\pm &= \pm 2\frac{\delta_r}{r_s}\sqrt{\omega^2_0-\omega^2},\\
\beta_\pm &= \pm \frac{{\rm 2}i}{\delta_r}\left[\frac{\omega}{r_s}\left(r_+\left(r_+-2D\right)+a^2\right)-m_\ell a\right], \label{beta}\\
\gamma_\pm &= \pm\frac{2i}{\delta_r}\left[\frac{\omega}{r_s}\left(r_-\left(r_--2D\right)+a^2\right)-m_\ell a\right] = \mp 2i K_3, \\
\delta &=-\frac{\delta_r}{r^2_s}\left(r_++r_--2D\right)\left[2\omega^2-\omega^2_0\right] = K_2 - K_4 - \frac{2\left(K_1 + K_3 \delta_r\right)\omega}{r_s},\\
\eta &= -K_2 + 2K_1\left(\frac{K_3}{\delta_r} + \frac{\omega}{r_s}\right).
\end{align}
After finding all of the confluent Heun's parameters, we can present the complete exact radial solution of the Klein-Gordon equation in the Kerr-EMDA black hole background as follows,
\begin{equation}
R=R_{0\pm} e^{-\frac{1}{2}\alpha_\pm y}y^{\frac{1}{2}\beta_\pm}{\left(y-1\right)}^{\frac{1}{2}\gamma} \operatorname{HeunC}\left(\alpha_\pm ,\beta_\pm ,\gamma_\pm ,\delta ,\eta ,-y\right), \label{radwave}
\end{equation}
where $R_{0\pm}$ is normalization constant. Therefore, the complete exact wave function is obtained as the following,
\begin{multline}
\psi = e^{-i\omega t}{S_\ell^{ m_\ell}}\left(\theta,\phi \right)e^{-\frac{1}{2}\alpha_\pm \left(\frac{r-r_+}{\delta_r}\right)}{\left(\frac{r-r_-}{\delta_r}\right)}^{\frac{1}{2}\gamma_\pm}\times\\ \left[R_{0\pm}{\left(\frac{r-r_+}{\delta_r}\right)}^{\frac{1}{2}\beta_\pm}\operatorname{HeunC}\left(\alpha_\pm ,\beta_\pm ,\gamma_\pm ,\delta ,\eta,-\frac{r-r_+}{\delta_r}\right)\right]. \label{radsol}
\end{multline}

\subsection{Energy Quantization}
The radial quantization condition is related to the termination condition of the radial wave function. The confluent Heun function will have $n_r$ zeros if the polynomial condition (see Appendix~\ref{AppendixB}) is fulfilled, 
\begin{gather}
\frac{\delta}{\alpha_\pm}+\frac{\beta_\pm +\gamma_\pm}{2} + 1 =-n_r,
\end{gather}
or explicitly,
\begin{multline}
\mp\frac{{\left(r_++r_--2D\right)}\left[2\omega^2-\omega^2_0\right]}{2r_s\sqrt{\omega^2_0-\omega^2}}
\\ \pm\frac{i}{\delta_r}\left\{\frac{\omega}{r_s}\left(r_+\left(r_+-2D\right)+r_-\left(r_--2D\right)+2a^2\right)-2m_\ell a\right\}=-n, \label{energylevelsmassive}
\end{multline}
where we have redefined a new radial quantum number as follows,
\begin{equation}
    n=n_r+1.
\end{equation}
In addition, one can also express the radial quantum number, $n$, in terms of total quantum number, $N$, as $N=n+\ell+1$ \cite{Yang,Yang22}.

Notice that the scalar energy levels have a dependence on the azimuthal quantum number $m_\ell$ that is directly coupled to the black hole's spin parameter $a$. This is similar with the Zeeman effect when a Hydrogenic atom is immersed in a magnetized space. The existence of the term can be understood as an interaction between the orbiting scalar field with magnetic state $m_\ell$, with the black hole's angular momentum $a$. 

\subsection{Quasibound States in Extreme Regions}
Now, we are going to investigate the behaviour of the exact quasibound states solution in two extreme regions, i.e., very near to the black hole's outer horizon, $r \to r_+$, and asymptotic behaviour far away from the black hole's horizon, $r \to \infty$. Since the quasibound states are quantized states, therefore the confluent Heun functions are always polynomial functions. 

Let us first consider how the quasibound states behave very near to the apparent horizon by taking the limit $r \to r_+$ in \eqref{energylevelsmassive}. In this limit the confluent Heun polynomial are approximately equal to unity. Also the exponential term is $e^{-\frac{1}{2}\alpha_{\pm} \left(\frac{r-r_+ }{\delta_r}\right)}\to 1$ as $r\to r_+$. 
The quasibound states exist in a semi-infinite potential well that causes the radial wavefunction vanishes at $r=\infty$. At the event horizon $r=r_+$, there is only ingoing mode. Now, let us impose the first boundary condition to the exact solution \eqref{radwave} by firstly rewriting the asymptotically far radial solution following the Appendix \ref{AppendixB},
\begin{equation}
   R(r\to \infty)\sim \frac{1}{r} e^{-\frac{1}{2}\alpha_\pm \left(\frac{r}{\delta_r}\right)}r^{-\frac{\delta}{\alpha_\pm}}.
\end{equation}
The condition $\omega_0^2-\omega^2>0$ implies that only $\alpha_+=+2 \frac{\delta_r}{r_s}\sqrt{\omega_0^2-\omega^2}$ satisfies the boundary condition at infinity. And since condition $\omega_0^2-\omega^2>0$ is assumed to be hold, only $\alpha_+=+2 \frac{\delta_r}{r_s}\sqrt{\omega_0^2-\omega^2}$ satisfies the boundary condition at $r=\infty$. 

The second boundary condition, at $r_+$, requires the solution to be purely ingoing, entering the the event horizon. Let us first rewrite the radial solution \eqref{radwave} in the near event horizon limit,
\begin{equation}
    R(r\to r_+) \sim \left(r-r_-\right)^{\frac{\beta_\pm}{2}},
\end{equation}
where $\frac{\beta_\pm}{2}$ can be written as follows,
\begin{gather}
   \frac{\beta_\pm}{2}=\pm\frac{i}{\kappa_+} \left(\omega-\omega _a\right),\\
   \kappa_+=\frac{\delta_r}{r_+(r_+-2D)+a^2}>0 \ \ \ , \ \ \ \omega _a=\frac{m_\ell a }{r_+(r_+-2D)+a^2}.
\end{gather}
We find only $\beta_-$ satisfies the boundary condition at the event horizon. Therefore, the temporal together with the radial part of the wavefunction at the horizon is
\begin{equation}
    T(t)R(r)\sim e^{-i\omega t}\left(r-r_-\right)^{-\frac{i}{\kappa_+} \left(\omega-\omega _a\right)}.
\end{equation}
Similar expression is obtained by doing series expansion of the radial solution in the near horizon limit of the Kerr-Newman black hole, where the radial solution is given in term of the two independent Whittaker confluent hypergeometric functions \cite{Rowan}. The same expression is obtained using purely asymptotical series method dilatonic charged black hole \cite{Yang}. We also find that only two out of eight modes satisfying the boundary conditions, i.e., the modes with $\alpha_+,\beta_-,\gamma_\pm$. One can also impose further restriction at the inner horizon by imposing ingoing wave condition, which eliminates the $\alpha_+,\beta_-,\gamma_+$ mode \cite{vier21}.

After imposing the boundary conditions, the exact formula of the massive scalar fields quantized eigenfrequencies, i.e., the mode $(\alpha_+, \beta_-,  \gamma_-)$ can be rewritten as follows,
\begin{multline}
-\frac{{\left(r_++r_--2D\right)}\left[2\omega^2-\omega^2_0\right]}{2r_s\sqrt{\omega^2_0-\omega^2}}
\\ -\frac{i}{\delta_r}\left\{\frac{\omega}{r_s}\left(r_+\left(r_+-2D\right)+r_-\left(r_--2D\right)+2a^2\right)-2m_\ell a\right\}=-n.\label{eigenfreq}
\end{multline}
For massless scalar fields, where $\omega_0=0$, we obtain this following expression,
{\begin{gather}
\omega_1=\frac{2m_\ell a-in(r_+-r_-)}{2\left(r_+(r_+-2D)+a^2\right)}r_s,\label{energylevelsmassless1}\\
\omega_2=\frac{2m_\ell a-in(r_+-r_-)}{2\left(r_-(r_--2D)+a^2\right)}r_s. \label{energylevelsmassless2}
\end{gather}}
It is also interesting to investigate the energy quantization expression \eqref{energylevelsmassive} for ultralight scalar fields, where $\omega_0^2<<1$ and $\omega_0^2-\omega^2<<1$. In this particular case, we obtain the following "gravitational atom" expression, 
\begin{align}
\omega&=\omega_0\sqrt{1-{\left[\frac{\omega_0{\left(r_++r_--2D\right)}}{2r_s\left(i\frac{2m_\ell a}{\delta_r}+n\right)}\right]}^2},\\
&\approx\omega_0\left(1 - \frac{\omega_0^2}{8 n^2}\right) + i\frac{\omega_0^3  m_\ell a}{2 n^3 \delta_r}+ \frac{3\omega_0^3 m_\ell^2 a^2}{2 n^4 \delta_r^2}.
\end{align}
which agrees with those obtained in \cite{Yang,Yang22}.


\subsection{Quasibound States Resonance Profiles}

\begin{figure}[h]
    \centering
    \includegraphics[scale=0.6]{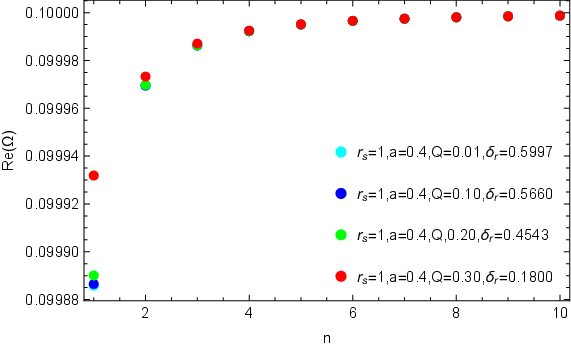}
    \includegraphics[scale=0.6]{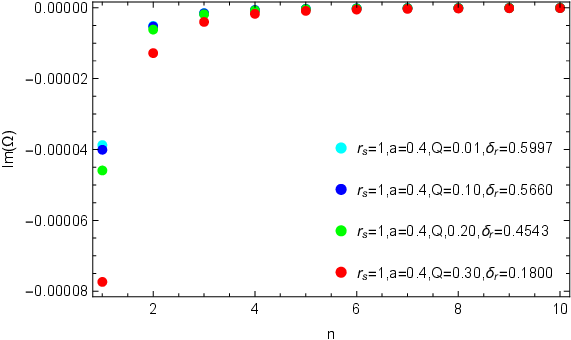}
    \caption{Selected quasibound states frequencies for several sub-extremal Kerr-EMDA black holes. For fixed $a=0.4,m_{\ell}=0$ and $\omega_0=0.1$. } \label{fig1}
\end{figure}

\begin{figure}[h]
    \centering
    \includegraphics[scale=0.6]{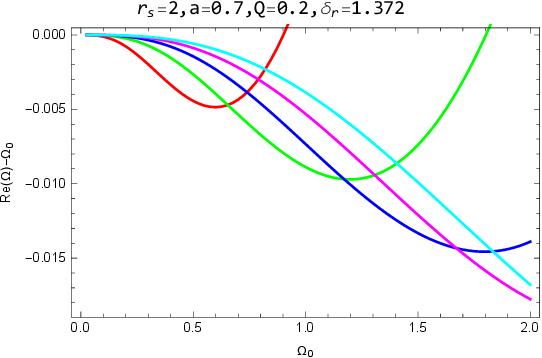}
    \includegraphics[scale=0.6]{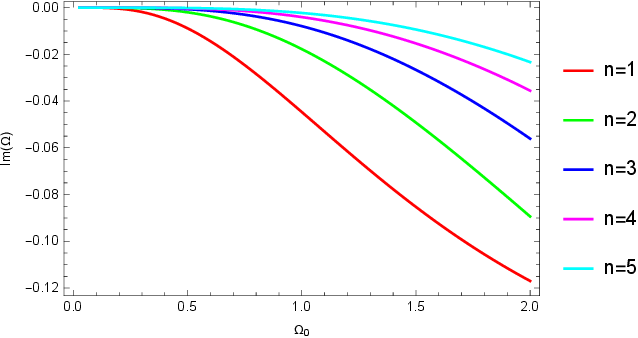}
    \caption{Quasibound states frequencies profiles of the spherical modes ($m_\ell=0$) for various scalar fields' masses.}
  \label{fig2}
\end{figure}

\begin{figure}[h]
    \centering
    \includegraphics[scale=0.6]{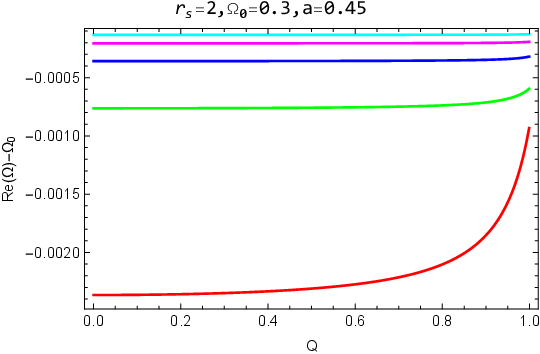}
    \includegraphics[scale=0.6]{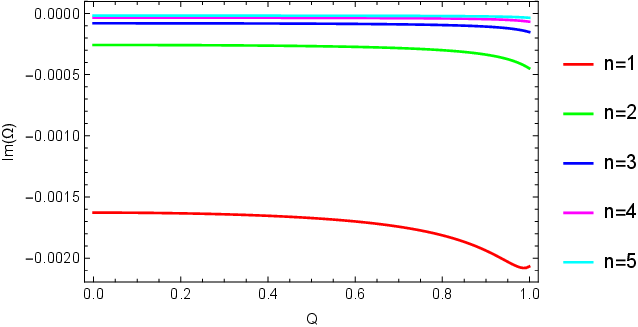}
    \caption{Quasibound states frequencies profiles of the spherical modes ($m_\ell=0$) for various black hole's charge.} 
  \label{fig3}
\end{figure}

\begin{figure}[H]
    \centering
    \includegraphics[scale=0.6]{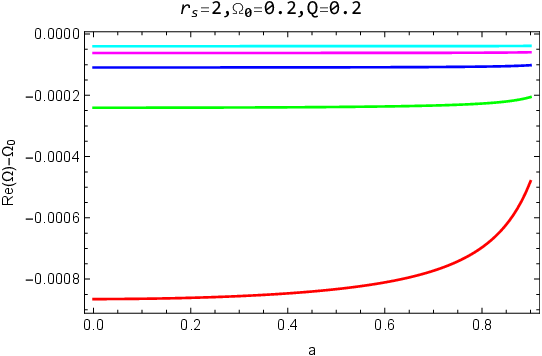}
    \includegraphics[scale=0.6]{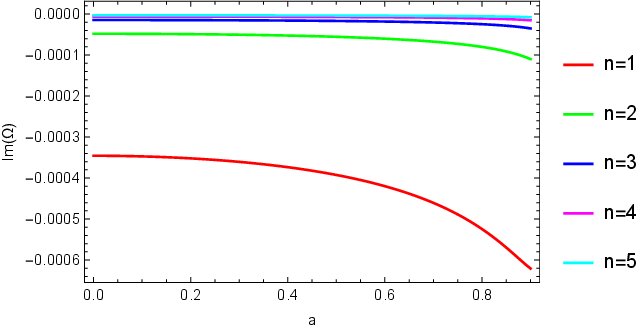}
    \caption{Quasibound states frequencies profiles of the spherical modes ($m_\ell=0$) for various black hole's spin. } 
  \label{fig4}
\end{figure}

Here in this subsection, we present QBS profiles of massive scalar field on sub-extremal Kerr-EMDA black holes. These are shown through \cref{fig1} -- \cref{fig4}.  We visualize  massive scalar's QBS frequencies in four different sub-extremal Kerr-EMDA black holes in \cref{fig1}. We plot the complex quasibound states frequencies against $n$ for several value of $\delta_r$. The scalar's mass is fixed to be $\omega_0=0.1$ the Kerr-EMDA black hole's mass is fixed such that $r_s=1$. As $n$ increases, both real and imaginary of $\omega$ decrease (in magnitude for $Im(\omega)$). As we approach extremal region, i.e., $\delta_r$ is smaller, the differences between the ground state and other excited states become less evident. 

In \cref{fig2,fig3,fig4}, we illustrate QBS frequencies for the first five radial quantum numbers $n=1, \dots ,5$ against scalar field's mass $\omega_0$, electric charge $D=-\frac{Q^2}{r_s}$ \cite{Garcia} and black hole's spin $a$ range from zero up to its near-extremal limit. In general, we notice that QBS frequency profiles exhibit strong shifts as one approaches the near-extremal states. In addition, the states characterized by lower radial quantum numbers, $n$, i.e., the low excited states, are more sensitive to the black hole's parameters changes compared to states with higher $n$.

\section{Scalar Cloud}\label{sect:cloud}
In this section, we will investigate the existence of the stationary scalar-black hole configuration known as the scalar cloud bound to the axisymmetric Kerr-EMDA black hole. The scalar cloud is bound state configuration of scalar field around the black hole where scalar field's frequency equals to superradiant threshold \cite{Benone:2014ssa}.

The neutral scalar clouds are made of bosons having purely real eigenfrequency $\omega$ that matches the black hole's event horizon angular velocity $\omega_a$. Mathematically the configuration is described as follows \cite{Garc,Ric,Herdeiro:2014goa},
\begin{align}
Re(\omega) &=\omega_a=\frac{m_\ell a }{r_+(r_+-2D)+a^2}>0, \label{cond0}\\
Im(\omega)&=0. \label{cond1}
\end{align}
The first scalar cloud condition is understood as the angular velocity resonance to the central spinning black hole. The second energy condition indicates that there is no leaking scalar flux impenetrates the outer horizon nor radiates to infinity. Therefore, they do not grow or decay over time. Both aforementioned conditions, \eqref{cond0} and \eqref{cond1}, need to be met to ensure the existence of the scalar cloud.

Let us now substitute the first condition of the scalar cloud \eqref{cond0} into the energy quantization condition \eqref{eigenfreq}. It is straightforward to obtain 
\begin{equation}
  -\frac{{r_s}\left[2\omega_a^2-\mu^2\right]}{2\sqrt{\mu^2-\omega_a^2}}+i \omega_a r_s=-n, \label{cloud}  
\end{equation}
where $\omega=\omega_a$. The analytical solution of the scalar cloud's eigenenergy equation \eqref{cloud} can be written in a series expansion as follows, 
\begin{multline}
\frac{\omega_a}{\mu}=\left[1-\frac{\mu^2r_s^2}{8n^2}+\frac{63\mu^4r_s^4}{128n^4}-\frac{2145\mu^6r_s^6}{1024n^6}+\frac{323323\mu^8r_s^8}{32768n^8}+O[\mu^{10}r_s^{10}]\right]\\+i\left[\frac{\mu^3r_s^3}{4n^3}-\frac{\mu^5r_s^5}{n^5}-\frac{9\mu^7r_s^7}{2n^7}-\frac{22\mu^9r_s^9}{n^9}+O[\mu^{11}r_s^{11}]\right].
\end{multline}
Notice that for light mass scalars, $\mu r_s<<1$, the expansion can be terminated up to $\mu^2 r_s^2$ and consequently, the imaginary part of the eigenfrequency can be omitted. There, we obtain
\begin{equation}
\frac{\omega_a}{\mu}\approx1-\frac{\mu^2r_s^2}{8n^2}.
\end{equation}
Remarkably, this limit is always imposed in each of analytical works by various authors done via asymptotic matching method \cite{Furuhashi,Hod:2013zza,Benone:2014ssa,Huang:2016qnk}. In \cite{Furuhashi}, the approximated analytical expression of $\omega$ is shown to be purely real and hydrogen-like energy expression for a scalar field in the Kerr-Newman spacetime in the limit $\omega r_s <<1 $. The similar formula also appears for the first time in the context of scalar cloud in \cite{Hod:2013zza}.

Now, let's suppose that the Kerr-EMDA black hole's parameters, i.e., mass, charge and spin, are known. One can determine the mass of the scalar field by solving $\mu(\omega_a)$ of the scalar cloud's energy equation \eqref{cloud}, consequently this following expression is obtained,
\begin{multline}
\left(\mu r_s \right)^2=\left[{\omega_a^2r_s^2}+\frac{\omega_a^4r_s^4}{4n^2}-\frac{7\omega_a^6r_s^6}{8n^4}+\frac{165\omega_a^8r_s^8}{64n^6}+O(\omega_a^{10}r_s^{10})\right]\\+i\left[-\frac{\omega_a^5r_s^5}{2n^3}+\frac{3\omega_a^7r_s^7}{2n^5}-\frac{143\omega_a^9r_s^9}{32n^7}+O(\omega_a^{11}r_s^{11})\right],
\end{multline}
where the imaginary part can consistently be omitted for light mass scalar fields,
\begin{equation}
\left(\mu r_s \right)^2\approx{\omega_a^2r_s^2}+\frac{\omega_a^4r_s^4}{4n^2}. \label{smallmucloud}
\end{equation}
\begin{figure}
    \centering
    \includegraphics[width=0.49\linewidth]{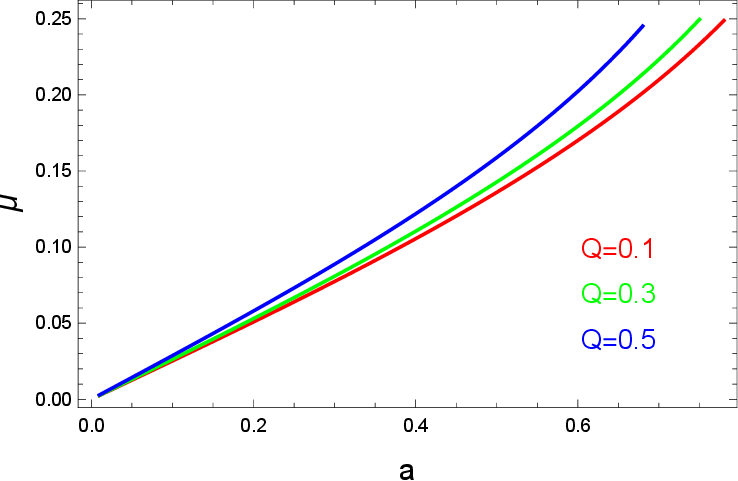}
    \includegraphics[width=0.49\linewidth]{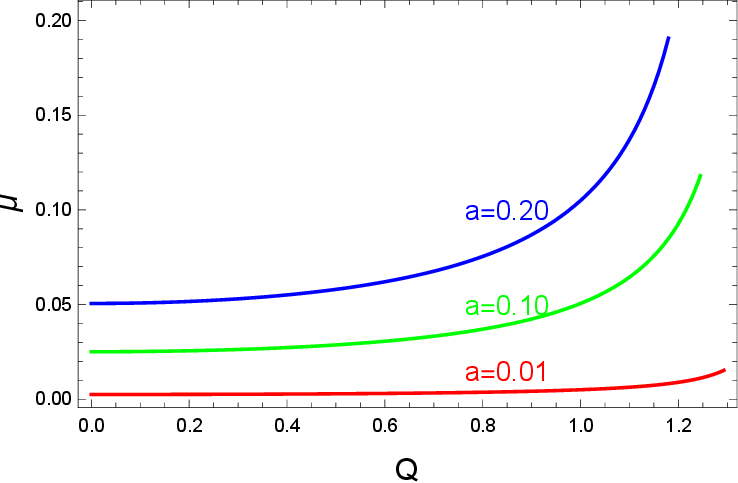}
    \caption{Scalar cloud curves for fixed $r_s=2,n=1,m_{\ell}=1$ Left: $Q=0.2$ and Right: $a=0.25$.}
    \label{fig:scalarcloud}
\end{figure}
In \cref{fig:scalarcloud}, we illustrate scalar cloud curves from \eqref{smallmucloud}. The relation between (light) scalar field's mass as a function of black hole's spin and charge are shown. As can be seen, the allowed scalar field's mass increases with spin and charge of the black holes.



\section{Boson Statistics Near Horizon}
\label{sect:Hawking}

In this section, we are going to investigate the boson statistics near the rotating EMDA black hole's event horizon. The scalar radiation distribution will be derived by utilizing the exact radial solutions revealing the black hole's superradiance condition. Let us start by writing the radial wave functions near the black hole's event horizon, $r\to r_+$, as follows,
\begin{gather}
R(r) \approx {\left(\frac{r_+-r_-}{\delta_r}\right)}^{\frac{1}{2}\gamma}\left[B{\left(\frac{r-r_+}{\delta_r}\right)}^{-\frac{1}{2}\beta_+}+A{\left(\frac{r-r_+}{\delta_r}\right)}^{\frac{1}{2}\beta_{+}}\right],
\end{gather}
where $A, B$ are constants and $\beta_+$ is given by \eqref{beta}. 
Notice that the radial wave can be decomposed as follows, 
\begin{align}
    R &\approx  \left\{
     \begin{aligned}
       & \psi_{+out}=A{\left(\frac{r-r_+}{\delta_r}\right)}^{\frac{1}{2}\beta_+} \hspace{0.2cm} \emph{,outgoing particle} \\
       & \psi_{+in}\hspace{0.15cm}=B{\left(\frac{r-r_+}{\delta_r}\right)}^{-\frac{1}{2}\beta_+}  \emph{,ingoing particle} 
     \end{aligned}
   \right. 
\end{align}
Following the Damour-Ruffini method \cite{Damour}, the unique expression for the absorbed antiparticle counterpart for each outgoing particle can be found via analytical continuation, i.e., $\psi_{-out}=\psi_{+out}\left(\left(\frac{r-r_+}{\delta_r}\right)\to \left(\frac{r-r_+}{\delta_r}\right)e^{-i\pi}\right)$. Thus, we obtain,
\begin{equation}
\begin{split}
\psi_{-out}&=A{\left(\left(\frac{r-r_+}{\delta_r}\right)e^{-i\pi}\right)}^{\frac{1}{2}\beta_+},\\
&=\psi_{+out}e^{-\frac{1}{2}i\pi \beta_+}. \label{anac}
\end{split}
\end{equation}
So, the relative probability amplitude to the ingoing wave is obtained as follows,
\begin{align}
{\left\lvert \frac{\psi_{-out}}{\psi_{+in}}\right\rvert }^2&={\left\lvert \frac{\psi_{+out}}{\psi_{+in}}\right\rvert }^2e^{-2i\pi \beta_+}={\left\lvert \frac{\psi_{+out}}{\psi_{+in}}\right\rvert }^2 e^\zeta, \label{analiticalcon}
\end{align}
where,
\begin{equation}
    \zeta =\frac{4\pi}{\delta_r}\left[\omega-\omega_a\right]\left(r_+\left(r_+-2D\right)+a^2\right).
\end{equation}
The distribution function of the emitted scalars is to be obtained via normalization condition of the total outgoing wave. First, let us consider the outgoing wave consisting particle and antiparticle wave. The wave function can be written with the help of the Heaviside step function as follows,
\begin{align}
\psi_{out}=\psi_{+out}\Theta \left(r-r_+\right)+\psi_{-out}\Theta \left(r_+-r\right).
\end{align}
The normalization condition leads to,
\begin{equation}
\left\langle{\left\lvert \frac{\psi_{out}}{\psi_{+in}}\right\rvert }^2\right\rangle = \left\langle{\left\lvert \frac{\psi_{+out}}{\psi_{+in}}\right\rvert }^2\right\rangle-\left\langle{\left\lvert \frac{\psi_{-out}}{\psi_{+in}}\right\rvert }^2\right\rangle=1,    
\end{equation}
and by using \eqref{analiticalcon} to eliminate ${\left\lvert \frac{\psi_{-out}}{\psi_{+in}}\right\rvert }^2$, it is straightforward to obtain this following bosonic distribution function,
\begin{gather}
\left\langle{\left\lvert \frac{\psi_{+out}}{\psi_{+in}}\right\rvert }^2\right\rangle=\frac{1}{\left\lvert 1-e^{\zeta}\right\rvert}. \label{bosondist}
\end{gather}
Whenever $\omega<{\omega_a}$, the distribution function \eqref{bosondist} grows exponentially indicating superradiant amplification. One could further analytically calculate the energy flux by the Hawking radiation as follows,
\begin{align}
\Phi_E &= \int_0^\infty \frac{\omega}{e^{\frac{4\pi}{\delta_r}\left[\omega-\omega_a\right]\left(r_+\left(r_+-2D\right)+a^2\right)}-1} d\omega, \\
&= \left[\frac{\delta_r}{4\pi\left(r_+\left(r_+-2D\right)+a^2\right)}\right]^2 Li_2\left(e^{\frac{4\pi m_\ell a}{\delta_r}}\right), \label{hawkingflux}
\end{align}
where the function $Li_n(x)$ is the Jonqui\'ere's Polylogarithm function \cite{NIST}. 

\begin{figure}[h]
    \centering
    \includegraphics[scale=1]{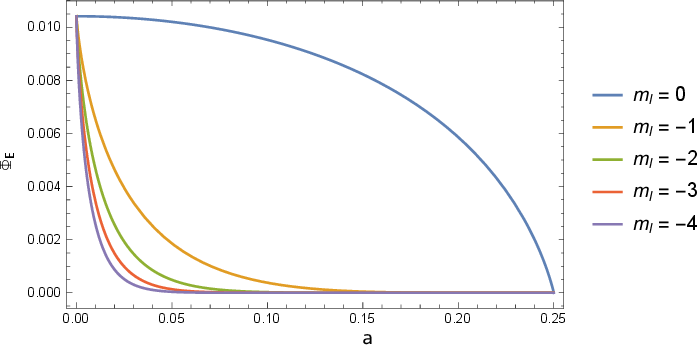}
    \caption{Hawking radiation flux profile as functions of the black hole's spin, $a$ for various $m_\ell$.} 
  \label{fig6}
\end{figure}

In \cref{fig6}, we show the Hawking radiation flux, $\Phi_E$, as functions of the black hole's spin, $a$. The black hole's parameters are set as follows: $r_s = 2$ and $Q = 0.5$ with $m_\ell$ is varied. We plot the radiation flux beginning at $a=0$ and extending to the extremal limit of the black hole, which is at $a=0.25$. Notice that the radiation flux approaches zero for all $m_\ell$ when the black hole reaches its extremal condition. Additionally, it is important to mention that positive $m_ \ell$ results in a complex-valued radiation flux, therefore it is discarded. 

\section{Superradiance and Wave Amplification}\label{sect:suprad}

In this section, we analytically derive the wave amplification coefficient of scalar fields around the Kerr-EMDA black hole via analytical asymptotic matching. The general idea of the asymptotic matching method is as follows: first, we have to work out the solutions in both of the two extreme limits, i.e., near the event horizon and at infinity, of which the solutions are expected to be obtained in terms of the hypergeometric functions. The connection formulas of the hypergeometric functions are then used to obtain the $r\to\infty$ limit from the near the event horizon solution. Both solutions are then matched in the intermediate region and the amplification coefficient is obtained. 

Let us rewrite the radial Klein-Gordon equation \eqref{radialmod} in this following form,
\begin{multline}
    \Delta {\partial }_r\left(\Delta {\partial }_rR\right)+\left[{\left\{\frac{\omega }{r_s}\left(r\left(r-2D\right)+a^2\right)-m_\ell a\right\}}^2\right. \\ \left.-\Delta \left\{\frac{{\omega }^2_0}{r^2_s}\left(r\left(r-2D\right)+a^2\right)+K^{m_\ell}_{\ell}\right\}\right]R=0.
\end{multline}
We define a new radial coordinate $x=\frac{r-r_+}{\delta_r}$. Therefore, black hole event horizon locates at $x=0$ and $x\to \infty$ as $r\to \infty$. In this section, we shall consider the radial equation only up to $a\omega$ term. Thus, the radial Klien-Gordon equation then takes the form,
\begin{align}
 x^2{\left(x+1\right)}^2{\partial }^2_xR+x\left(x+1\right)\left(2x+1\right){\partial }_xR+\left[H(x)+I(x)\right]R&=0,\label{radialx} 
\end{align}
where,
\begin{align}
H(x)&=\left(\frac{\omega }{\delta_rr_s}\left(\left(\delta_rx+r_+\right)\left(\delta_rx+r_+-2D\right)\right)-\frac{m_\ell a}{\delta_r}\right)^2, \\
I(x)&=-x\left(x+1\right)\left(\frac{{\omega }^2_0}{r^2_s}\left(\left(\delta_r x+r_+\right)\left(\delta_rx+r_+-2D\right)\right)+\lambda^{m_\ell}_\ell -2\frac{m_\ell a}{r_s}\omega\right).
\end{align}
To proceed further, we introduce two short-hand notation,
\begin{align}
\left(\delta_rx+r_+\right)\left(\delta_rx+r_+-2D\right)&\equiv X(r) + r_+(r_+-2D), \nonumber \\
\frac{\Delta_\omega}{\delta_r r_s}&\equiv\left(\frac{\omega }{\delta_rr_s}-\frac{m_\ell a}{\delta_r r_+(r_+-2D)}\right)r_+(r_+-2D).\label{DeltaOhm}
\end{align}
With these, we can write $H(x)$ in a more compact form,
\begin{align}
H(x)&=\left(\frac{\omega }{\delta_rr_s}X+ \frac{\Delta_\omega}{\delta_r r_s} \right)^2, \nonumber \\
&=\frac{\omega^2}{\delta_r^2r_s^2}\left[\delta_r^4 x^2(x+1)^2+2\delta_r^3 r_s x^2(x+1)+\delta_r^2 r_s^2 x^2\right] \nonumber \\
&~~~~~+2\frac{\omega \Delta_\omega}{\delta_r^2r_s^2}\left[\delta_r^2 x(x+1)+\delta_r r_s x\right]+\frac{\Delta_\omega^2}{\delta_r^2 r_s^2}. \label{Hx}
\end{align}
Now, let us consider $I(x)$. We modify it in the similar manner as follows,
\begin{align}
    I(x)&=-x\left(x+1\right)\Bigg(\frac{{\omega }^2_0}{r^2_s}\delta_r^2 x(x+1) +\frac{{\omega }^2_0}{r^2_s}\delta_r r_s x +  \frac{\omega_0^2}{r_s^2}r_+\left(r_+-2D\right) \nonumber \\
    &~~~~~+\lambda^{m_\ell}_\ell -2\frac{m_\ell a}{r_s}\omega\Bigg). \label{Ix}
\end{align}
As $x\to\infty$, $H$ and $I$ can be obtained as follows,
\begin{align}
    H(x) &\approx \frac{\delta_r^2\omega^2}{r_s^2}x^4 + \frac{4(r_+-D)\delta_r\omega^2}{r_s^2}x^3 \nonumber \\
    &~~~- \frac{2\omega\left(\omega\left(6Dr_+-3r_+^2 - 2D^2 \right)+ a m r_s\right)}{r_s^2}x^2, \\
    I(x) &\approx -\frac{\delta_r^2\omega_0^2}{r_s^2}x^4 + \frac{\left(2D-2r_+-\delta_r\right)\delta_r \omega_0^2}{r_s^2}x^3 \nonumber \\
    &~~~+ \left(\frac{2am\omega r_s + \left(2D\left(r_+ + \delta_r\right)-r_+\left(r_++2\delta_r\right)\right)\omega_0^2}{r_s^2}-\lambda_\ell^{m_\ell}\right)x^2.
\end{align}
The far region radial equation becomes,
\begin{align}
   0 &= x^2 R'' + 2x R' + \left[Ax^2 +Bx+ C\right]R, \label{farfieldequation1} \\
   A &= \frac{\left(\omega^2-\omega_0^2\right)\delta_r^2}{r_s^2}, \\
   B &= \frac{2\delta_r\left(r_s+\delta_r\right)}{r_s^2} \omega^2 - \frac{\delta_r\left(r_s+2\delta_r\right)}{r_s^2}\omega_0^2, \\
   C &= -\lambda_{\ell}^{m_{\ell}} + \frac{2\left(3r_+^2-6Dr_++2D^2\right)}{r_s^2}\omega^2 + \frac{\left( 
  2D\left(r_++\delta_r\right)-r_+\left(r_++2\delta_r\right)\right)}{r_s^2}\omega_0^2. \label{Cconstant}
\end{align}
By following the Appendix \ref{AppendixA}, the normal form of the far field equation \eqref{farfieldequation1} can be obtained as follows, 
\begin{gather}
    \mathcal{R}''+\left[A+\frac{B}{x}+\frac{C}{x^2}\right]\mathcal{R}=0,
\end{gather}
where $\mathcal{R}=x R$. By redefining a new radial variable,  $z=2i\sqrt{A} x$, we obtain the following equation,
\begin{gather}
\mathcal{R}''+\left[-\frac{1}{4}+\frac{B}{2i\sqrt{A}z}+\frac{C}{z^2}\right]\mathcal{R}=0. \label{normalformfarfieldequation1}
\end{gather}
The general solution to \eqref{normalformfarfieldequation1} can be written in terms of the confluent hypergeometric functions $F_{1,1}\left(a,b,x\right)$ as follows (see Appendix \ref{AppendixC}),
\begin{align}
R_{far}&=N_{\infty 1}e^{-\frac{1}{2}i\alpha x}x^{\beta-\frac{1}{2}}F_{1,1}\left(\frac{1}{2}+\beta+i\frac{B}{\alpha}, 1+2\beta,\ i\alpha x\right) \nonumber \\
&~~~~+N_{\infty 2}e^{-\frac{1}{2}i\alpha x}x^{-\beta-\frac{1}{2}}F_{1,1}\left(\frac{1}{2}-\beta+i\frac{B}{\alpha},\ 1-2\beta,\ i\alpha x\right), \label{farsol}
\end{align}
where,
\begin{align}
    \alpha&=2\sqrt{A},~~~~~~\beta=\frac{1}{2}\sqrt{1-4C}.
\end{align}
The arbitrary constants are given by $N_{\infty 1}$ and $N_{\infty 2}$. Finally, we consider the $x\to 0$ limit. The near horizon solution can be obtained as, 
\begin{equation}
R_{far}\left(x\to 0\right)\approx N_{\infty 1}x^{\beta-\frac{1}{2}}+N_{\infty 2}x^{-\beta-\frac{1}{2}},
\end{equation}
where $F_{1,1}\left(a,b,0\right)=1$ is applied. 

Let us now turn to derive the solutions to the radial equation \eqref{radialx} close to the black hole's event horizon i.e., $x\to 0$. Let us first re-write $\frac{H(x)}{x(x+1)}$ as follows,
\begin{gather}
\frac{H(x)}{x(x+1)}=\frac{\delta_r^2 \omega^2}{r_s^2}x^2+\frac{\delta_r(\delta_r+2r_s)\omega^2}{r_s^2}x+\omega^2+2\frac{\Delta_\omega\omega}{r_s^2}+\frac{\Delta_\omega^2}{\delta_r^2r_s^2x}-\frac{\left(\Delta_\omega-\delta_rr_s\omega\right)^2}{\delta^2r_s^2(x+1)},
\end{gather}
where in $x\to 0$ limit, we drop the first two terms and obtain,
\begin{equation}
 \frac{H(x)}{x(x+1)}\approx\omega^2+2\frac{\Delta_\omega\omega}{r_s^2}+\frac{\Delta_\omega^2}{\delta_r^2r_s^2x}-\frac{\left(\Delta_\omega-\delta_rr_s\omega\right)^2}{\delta^2r_s^2(x+1)}.   
\end{equation}
Thus, the radial equation \eqref{radialx} in the limit $x\to 0$ takes the form,
\begin{equation}
x\left(x+1\right)R''+\left(2x+1\right)R'+\left[\frac{1}{x}\frac{\Delta_\omega^2}{\delta_r^2 r_s^2}-\frac{1}{(x+1)}\frac{\left(\Delta_\omega-\delta_rr_s\omega\right)^2}{\delta_r^2r_s^2}+G\right]R=0, \label{10}
\end{equation}
where we have defined a dimensionless constant,
\begin{gather}
G\equiv-\left(\lambda^{m_\ell}_\ell-\omega^2+\frac{\omega_0^2}{r_s^2}r_+\left(r_+-2D\right)-2\frac{m_\ell a}{r_s}\omega-2\frac{\Delta_\omega\omega}{r_s^2}\right).
\end{gather}
We notice that $G$ has similar structure as $C$ previously defined in \eqref{Cconstant}. In fact, they agree up to first order in $\omega$ and $\omega_0$ i.e., $C-G = O(\omega^2,\omega_0^2)\mathcal{}$. This condition can also be understood as limiting the calculation only for slowly rotating black hole and for bosons having much larger Compton wavelength comparing to the rotating black hole's size. 

\begin{figure}[h]
    \centering
\includegraphics[scale=0.35]{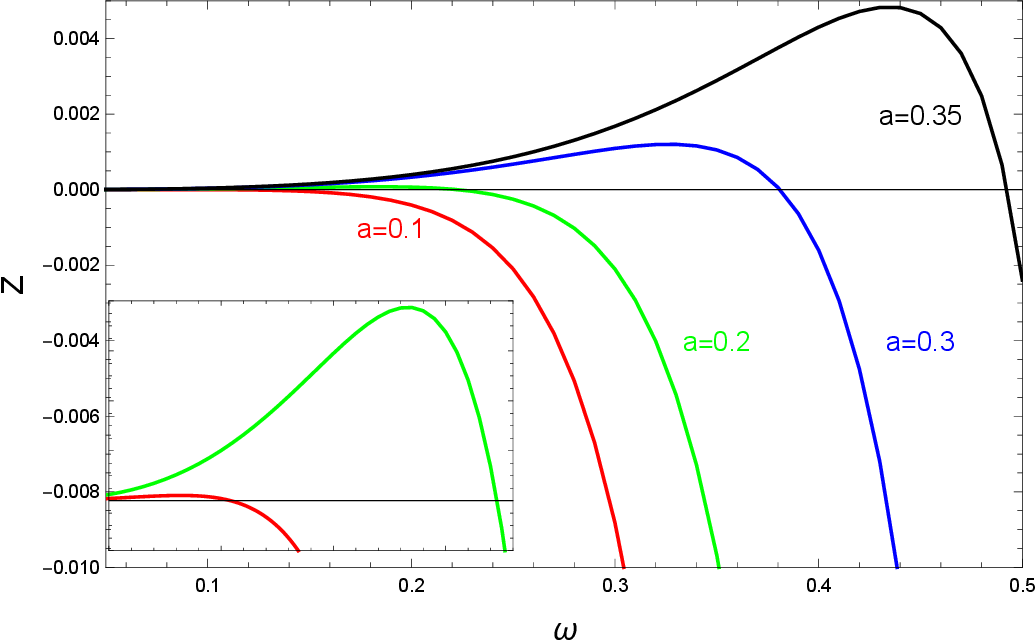}
\includegraphics[scale=0.35]{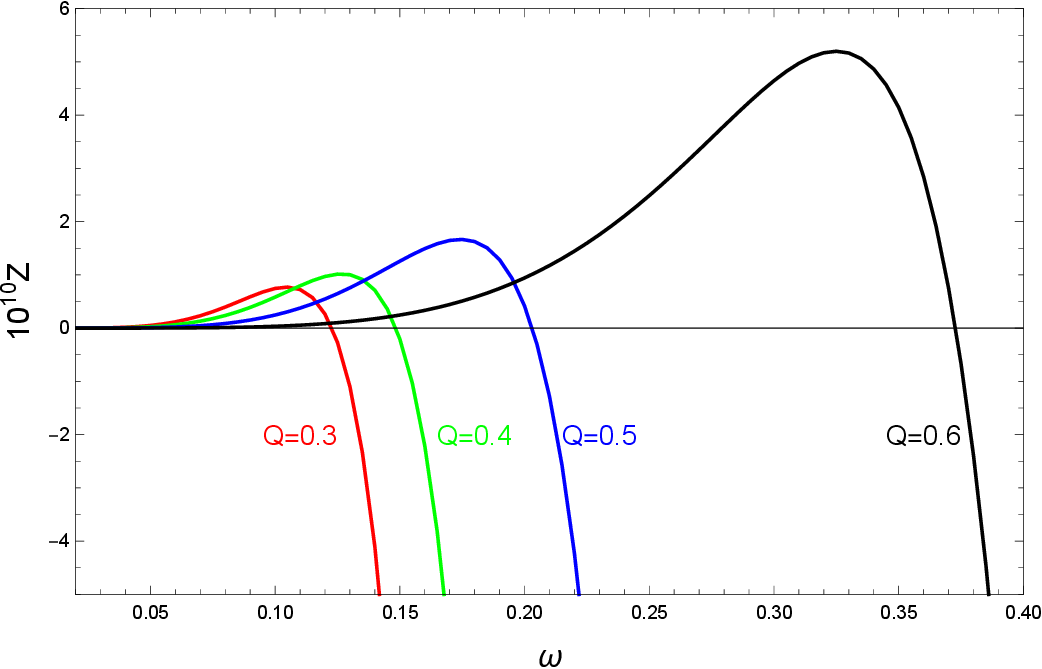}
    \caption{The amplification factor $Z_{\ell,m_\ell,a}$ as a function of $\omega$ for $r_s=1$ and $\omega_0=0.01$. Left: For $Q=0.1,\ell=1$ and $m_{\ell}=1$. The sub-figure display behaviour at small $\omega$.  Right: For $a=0.05, \ell=2$ and $m_{\ell}=2$. In this plot, the amplification factor is multiplied by factor of $10^{10}$. } 
  \label{fig:normalscale}
\end{figure}

\begin{figure}[h]
    \centering
    \includegraphics[scale=0.9]{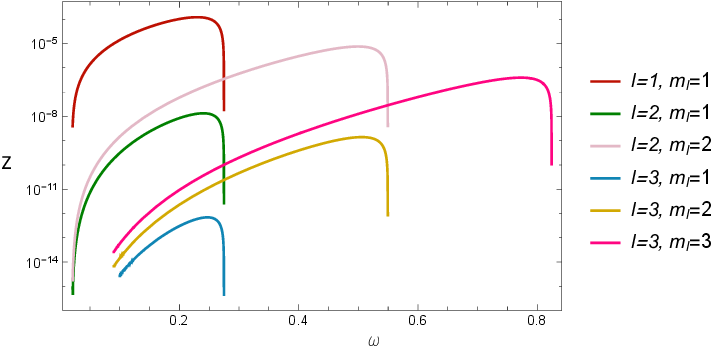}
    \caption{The $ Z_{\ell,m_\ell,a}$ profile  for various $\ell,m_\ell$, and fixed $r_s=1, a=0.2, Q=0.3, \omega_0=0.02$. The plot is displayed in semi-$\log$ graph.} 
  \label{Amplification}
\end{figure}

\begin{figure}[H]
    \centering
    \includegraphics[scale=0.9]{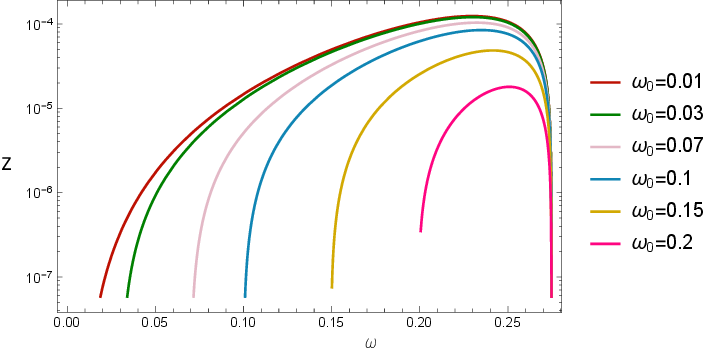}
    \caption{$ Z_{\ell,m_\ell,a}$ profile  for various $\omega_0$, and fixed $\ell=1, m_\ell=1, r_s=1, a=0.2, Q=0.3$.} 
  \label{Amplification1}
\end{figure}

\begin{figure}[H]
    \centering
    \includegraphics[scale=0.9]{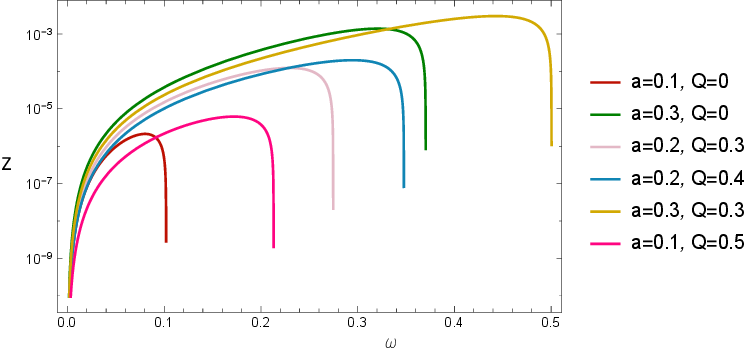}
    \caption{$ Z_{\ell,m_\ell,a}$ profile  for various $a, Q$, and fixed $\ell=1, m_\ell=1, r_s=1,\omega_0=0.001$.} 
  \label{Amplification2}
\end{figure}

Comparing the near horizon equation \eqref{10} with \eqref{gauss1}, we obtain the following ingoing wave solution,
\begin{equation}
       R_{near}=N_{H1}(x+1)^{-i\left(\frac{\Delta_\omega-\delta_rr_s\omega}{\delta_r r_s}\right)}x^{-\frac{i\Delta_\omega}{\delta_r r_s}}F_{2,1}\left(a_1,a_2,a_3,-x\right),
\end{equation}
where,
\begin{align}
    a_1&=\frac{1}{2}+\frac{1}{2}\sqrt{1-4G}-i\left(\frac{2\Delta_\omega-\delta_rr_s\omega}{\delta_r r_s}\right),\\
    a_2&=\frac{1}{2}-\frac{1}{2}\sqrt{1-4G}-i\left(\frac{2\Delta_\omega-\delta_rr_s\omega}{\delta_r r_s}\right),\\
    a_3&=1-2i\frac{\Delta_\omega}{\delta_r r_s}.
\end{align}
We utilize the identity \eqref{connection} to represent the solution near the black hole's horizon in terms of $ \frac { 1 } { x } $. This enables us to explore the limit as $ x \to \infty $ as follows,
\begin{equation}
   R_{near}\left(x\to \infty \right)\approx N_{H1}x^{-i\left(\frac{2\Delta_\omega-\delta_rr_s\omega}{\delta_r r_s}\right)}\left[x^{-a_1}\frac{\Gamma \left(a_3\right)\Gamma \left(a_2-a_1\right)}{\Gamma \left(a_3-a_1\right)\Gamma \left(a_2\right)}+x^{-a_2}\frac{\Gamma \left(a_3\right)\Gamma \left(a_1-a_2\right)}{\Gamma \left(a_3-a_2\right)\Gamma \left(a_1\right)}\right]. 
\end{equation}
In the overlap region, i.e., $R_{far}\left(x\to 0\right)=R_{near}\left(x\to \infty \right)$ the solutions are matched as follows,
By applying $C-G\to 0$ limit, we obtain
\begin{align}
-i\left(\frac{2\Delta_\omega-\delta_rr_s\omega}{\delta_r r_s}\right)-a_1 &= -\beta-\frac{1}{2},\\
-i\left(\frac{2\Delta_\omega-\delta_rr_s\omega}{\delta_r r_s}\right)-a_2 &= \beta-\frac{1}{2},
\end{align}
Therefore, we conclude that,
\begin{gather}
N_{\infty 1}=N_{H1}\frac{\Gamma \left(a_3\right)\Gamma \left(a_1-a_2\right)}{\Gamma \left(a_3-a_2\right)\Gamma \left(a_1\right)},\\
N_{\infty 2}=N_{H1}\frac{\Gamma \left(a_3\right)\Gamma \left(a_2-a_1\right)}{\Gamma \left(a_3-a_1\right)\Gamma \left(a_2\right)}.
\end{gather}
The quantity $|R|^2\equiv {\left|\frac{N_{\infty 1}}{N_{\infty 2}}\right|}^2$ is known as the reflection coefficient. In addition, we wish to obtain the amplification factor which is important when discussing superradiant phenomena. To do so, we have to express the far field solutions \eqref{farsol} in their asymptotic far limit. We follow asymptotic formula for $F_{1,1}(a,b,|x|\to \infty)$ as defined in \eqref{faratfar}. It is straightforward to obtain the asymptotical expansion of the first part of the far field solution of \eqref{farsol} follows, 
\begin{multline}
N_{\infty 1}e^{-\frac{1}{2}i\alpha x}x^{\beta-\frac{1}{2}}F_{1,1}\left(\frac{1}{2}+\beta+i\frac{B}{\alpha}, 1+2\beta,\ i\alpha x\right) = \nonumber \\
N_{\infty 1}\Bigg[(i\alpha)^{-\frac{1}{2}-\beta+\frac{iB}{\alpha}}x^{\frac{iB}{\alpha}-1}e^{\frac{i\alpha}{2}x}\frac{\Gamma\left(1+2\beta\right)}{\Gamma\left(\frac{1}{2}+\beta+\frac{iB}{\alpha}\right)}  \nonumber \\   
+ (-i\alpha)^{-\frac{1}{2}-\beta-\frac{iB}{\alpha}}x^{-\frac{iB}{\alpha}-1}e^{-\frac{i\alpha}{2}x}\frac{\Gamma\left(1+2\beta\right)}{\Gamma\left(\frac{1}{2}+\beta-\frac{iB}{\alpha}\right)} \Bigg],
\end{multline}
while the second part of the solution reads,
\begin{multline}
N_{\infty 2}e^{-\frac{1}{2}i\alpha x}x^{-\beta-\frac{1}{2}}F_{1,1}\left(\frac{1}{2}-\beta+i\frac{B}{\alpha},\ 1-2\beta,\ i\alpha x\right)=\\
    N_{\infty 2}\Bigg[(i\alpha)^{-\frac{1}{2}+\beta+\frac{iB}{\alpha}}x^{\frac{iB}{\alpha}-1}e^{\frac{i\alpha}{2}x}\frac{\Gamma\left(1-2\beta\right)}{\Gamma\left(\frac{1}{2}-\beta+\frac{iB}{\alpha}\right)} 
    \\+(-i\alpha)^{-\frac{1}{2}+\beta-\frac{iB}{\alpha}} x^{-\frac{iB}{\alpha}-1}e^{-\frac{i\alpha}{2}x}\frac{\Gamma\left(1-2\beta\right)}{\Gamma\left(\frac{1}{2}-\beta-\frac{iB}{\alpha}\right)}\Bigg].
\end{multline} 
Now, we can express the asymptotically far, far field solutions, as superposition of ingoing and outgoing waves as follows,
\begin{align}
  R_{far}(x\to \infty)&= A_{in} e^{-\frac{1}{2}i\alpha x} x^{-\frac{iB}{\alpha}-1} +A_{out} e^{\frac{1}{2}i\alpha x} x^{\frac{iB}{\alpha}-1},
\end{align}
with,
\begin{align}
A_{out} &\equiv N_{\infty 1}(i\alpha)^{-\frac{1}{2}-\beta+\frac{iB}{\alpha}}\frac{\Gamma\left(1+2\beta\right)}{\Gamma\left(\frac{1}{2}+\beta+\frac{iB}{\alpha}\right)}+N_{\infty 2}(i\alpha)^{-\frac{1}{2}+\beta+\frac{iB}{\alpha}}\frac{\Gamma\left(1-2\beta\right)}{\Gamma\left(\frac{1}{2}-\beta+\frac{iB}{\alpha}\right)},\\
A_{in}&\equiv N_{\infty 1}(-i\alpha)^{-\frac{1}{2}-\beta-\frac{iB}{\alpha}}\frac{\Gamma\left(1+2\beta\right)}{\Gamma\left(\frac{1}{2}+\beta-\frac{iB}{\alpha}\right)}+N_{\infty 2}(-i\alpha)^{-\frac{1}{2}+\beta-\frac{iB}{\alpha}}\frac{\Gamma\left(1-2\beta\right)}{\Gamma\left(\frac{1}{2}-\beta-\frac{iB}{\alpha}\right)}.  
\end{align}
Using the explicit expressions of $A_{in}$ and $A_{out}$, we can calculate the amplification factor, which is defined as follows \cite{refcoef,Suphot,Brito},
\begin{align}
    Z_{\ell,m_\ell,a}&\equiv {\left|\frac{A_{out}}{A_{in}}\right|}^2-1. \label{SuperAmp}
\end{align}
The amplification factor $ Z_{\ell,m_\ell,a}$ has physical meaning as follows: $ Z_{\ell,m_\ell,a}>0$ shows a gain in amplification factor and indicates a superradiance while the case with $ Z_{\ell,m_\ell,a}<0$ indicates a loss in amplification, corresponding to the non-existence of the superradiance. The threshold of superradiance i.e., $\omega=\omega_a$ (see \eqref{cond0}) marks the location where $Z_{\ell,m_\ell,a}(\omega_a)=0$. Now, we are ready to explore the behaviour of the amplification factor $Z_{\ell,m_\ell,a}$ \eqref{SuperAmp} in term of $\omega$. We emphasize that all the cases considered in Fig.~\ref{fig:normalscale}--\ref{Amplification2} are chosen to ensure that all the black holes are sub-extremal. Moreover, we need to ensure that $a\omega << 1$ to satisfy with our approximation mentioned earlier. 

In Fig.~\ref{fig:normalscale}, we explore the behaviour of the amplification factor $Z_{\ell,m_\ell,a}$ \eqref{SuperAmp} in term of $\omega$. With other parameters fixed, it is clear that increasing in spin parameter $a$ and $Q$ (which corresponding to dilaton parameter $D$) increase maximum value of $Z_{\ell,m_\ell,a}$. In this figure, we notice that $Z_{\ell,m_\ell,a}$ can be positive, zero and negative. By numerical investigation, we find that $Z_{\ell,m_\ell,a}$ indeed vanishes at the threshold of superradiant condition i.e., $Z(\omega_a)=0$. We remark the effect of spin parameter $a$ on $Z_{\ell,m_\ell,a}$ shares similar behaviour of the Kerr and Kerr-like black holes \cite{Brito,Franzin}.

In Fig.~\ref{Amplification}, we plot the amplification factor $ Z_{\ell,m_\ell,a}$ for various combinations of $\ell,m_\ell$ with fixed black hole parameters $r_s=1, a=0.2, Q=0.3$ and scalar mass $\omega_0=0.02$. The numerical superradiance cutoff frequency, where the amplification factor is zero or where the curves drop steeply in the Fig.~\ref{Amplification}, corresponds to the analytical formula in equation \eqref{cond0}, i.e., at $\omega_a= 0.2604 m_\ell$, which is a linear function of $m_\ell$. We remark that, we also explore the amplification factor for  negative $m_\ell$ modes which resulting in no superradiance effect are found i.e., $ Z_{\ell,m_\ell,a} < 0$. This is no surprising since superradiant condition for rotating black hole with a presence of massive scalar field is $\omega_0<\omega<\omega_a$ \cite{Khodadi:2020cht,Franzin}.

The effect of scalar field mass on $ Z_{\ell,m_\ell,a}$ is investigated in Fig.~\ref{Amplification1} for fixed $\ell=1, m_\ell=1, r_s=1, a=0.2, Q=0.3$. Unlike quasibound states and scalar clouds, superradiant states are not bound where $\omega_0 < \omega$  For a fixed $\omega_a$, scalar fields with lighter masses have a broader frequency range in the superradiant regime, i.e., $\omega_0<\omega<\omega_a$. It is also worth noting that the cutoff frequency is determined solely by the black hole's properties and not by the scalar mass.

In Fig.~\ref{Amplification2}, we set $r_s=1,\omega_0=0.001$ and plot various massive scalars amplification factor for modes $\ell=1,m_\ell=1$ by varying the black hole's parameters. We observe that increasing $a$ and $Q$ results in a higher cut-off frequency. 

\section{Conclusions}\label{sect:conclud}
In this work, we investigate  relativistic scalar fields bound to a rotating black hole solution of the Einstein-Maxwell-Dilaton-Axion theory. The relativistic field equation is governed by the Klein-Gordon equation. We have successfully solved the Klein-Gordon equation by using the separation of variable ansatz and obatining the exact solutions of the angular part in terms of Spheroidal Harmonics. 

The radial equation is then put into its normal form. We find that the general solution of the radial equation can be presented as the confluent Heun differential equation. To ensure the regularity of the radial equation at infinity, we implement the polynomial condition of the confluent Heun function. As a result, we obtain quantization of energy \eqref{energylevelsmassive}. In low energy limit, we obtain the $n^{-2}$ expression which is the so called "gravitational atom" \cite{Daniel, Anal1, Anal2, Anal3, Anal10, David-Senjaya}. For the special massless scalar case, we obtain a purely imaginary energy levels \cref{energylevelsmassless1,energylevelsmassless2}. 

We continue our investigation into a specific form of bound state known as the scalar cloud, where the boson eigenfrequency $\omega$ matches the black hole’s event horizon angular velocity $\omega_a$. We find that such configuration is possible around the Kerr-EMDA black hole for light-mass scalar fields. We successfully derive an analytical expression of the scalar cloud's energy condition, presented in equation \eqref{smallmucloud}.

The radiation distribution function \eqref{bosondist} is successfully derived in the area around the external black hole's horizon by expanding the exact radial solution to the first linear order of $r-r_+$. We find that the distribution is Bose-Einsteinian when $\omega\geq \omega_a$ and when $\omega<\omega_a$, the distribution function blows up exponentially, indicating the existence of superradiant amplification. We further investigate the Hawking radiation by integrating the obtained distribution function. We get the energy flux expression in terms of the Jonqui\'ere's Polylogarithm function \eqref{hawkingflux} and show its profile in Fig.~\ref{fig6}. The radiation flux profile indicates that states with lower $m_\ell$ are more dominant. Adjusting the Kerr-EMDA spin parameter to the extreme limit results in a large flux loss for states with non-zero $m_\ell$. Higher $m_\ell$ states have a lower probability of escaping the black hole with higher $a$, as seen by the steeper flux decrease. The Hawking radiation flux converges to zero for all modes at the extremal point, coinciding with the well-known zero temperature of the extremal black hole. 

In the last section, we calculate the amplification factor $Z_{\ell,m_\ell,a}$ via the asymptotic matching technique. 
We successfully obtain the analytically approximate amplification factor  \eqref{SuperAmp} and visualize the amplification factor for various scenarios. We find the amplification factor becomes positive whenever the condition $\omega_0 < \omega < \omega_a$ is fulfilled, which means for a given incoming wave with positive $m_\ell$ approaching a rotating black hole's event horizon, the reflected wave will be superradiantly amplified. We also find that the amplification factor reach higher maximum value with larger value of black hole's spin parameter $a$ and dilaton parameter $D$.

As a general extension of this work, obtaining an exact formula or semi-analytic formula for quasinormal modes and greybody factor of a rotating black hole in the EMDA theory can prove to be useful. We leave this investigation as future work.


\section*{Acknowledgments}
David Senjaya acknowledges this research project is supported by the Second Century Fund (C2F), Chulalongkorn University. SP acknowledges funding support from the NSRF via the Program Management Unit for Human Resources \& Institutional Development, Research and Innovation [grant number B39G670016].

\begin{appendices} 
    \include{appendix} 
\section{Normal Form} \label{AppendixA}
The so called "Normal Form" of an ordinary differential equation is the form when an ordinary differential equation is solved explicitly for the highest derivative \cite{NIST}. One may start with a general form of a linear second order ordinary differential equation as follows,
\begin{equation}
    \frac{d^2y}{dx^2}+p(x)\frac{dy}{dx}+q(x)y=0. \label{general ODE}
\end{equation}
We continue with applying a homotopic-family transformation, i.e., by making a special form of substitution for $y(x)$, which aims to remove the first order derivative term as follows \cite{2420},
\begin{align}
y&=Y(x)e^{-\frac{1}{2}\int{p(x)}dx},\\
\frac{dy}{dx}&=\frac{dY}{dx}e^{-\frac{1}{2}\int{p(x)}dx}-\frac{1}{2}Ype^{-\int{p(x)}dx},\\
\frac{d^2y}{dx^2}&=\frac{d^2Y}{dx^2}e^{-\frac{1}{2}\int{p(x)}dx}-\frac{1}{2}\frac{dY}{dx} pe^{-\frac{1}{2}\int{p(x)}dx}\nonumber \\
&~~~~-\frac{1}{2}Y\frac{dp}{dx}e^{-\frac{1}{2}\int{p(x)}dx}+\frac{1}{4}Yp^2e^{-\frac{1}{2}\int{p(x)}dx}.
\end{align}
After substituting the expressions to \eqref{general ODE}, we obtain second order differential equation without the first order derivative,
  \label{Normal Form}
\begin{gather}
    \frac{d^2Y}{dx^2}+\left(-\frac{1}{2}\frac{dp}{dx}-\frac{1}{4}p^2+q\right)Y=0. \label{normalform}
\end{gather}
The normal form is extremely useful to understand and estimate the physical nature of the ordinary differential equation. Suppose $Q(x)=-\frac{1}{2}\frac{dp}{dx}-\frac{1}{4}p^2+q>0$, the normal form reads $\frac{d^2Y}{dx^2}=-Q(x)Y$. Suppose we start with the case where $Y$ is positive, then, the second derivative will be negative and vice versa. It is clear that $Y$ will cross $x$-axis. If $Q(x)>0$ and $\int_1^\infty Q(x) dx=\infty$, then $Y(x)$ has infinitely many zeros on the positive $x$-axis. If $Q(x)<0$ then $Y(x)$ does not oscillate at all and has at most one zero  \cite{2420}.

\section{The Confluent Heun Equation and Its Solutions}\label{AppendixB}
The confluent Heun differential equation is a linear second order ordinary differential equation having the following canonical form \cite{Heun},  
\begin{align}
    \frac{d^2\psi_H}{dx^2}+\left(\alpha +\frac{\beta +1}{x}+\frac{\gamma +1}{x-1}\right)\frac{d\psi_H}{dx}+\left(\frac{\mu }{x}+\frac{\nu }{x-1}\right)\psi_H=0,
\end{align}
where,
\begin{align}
\mu&=\frac{1}{2}\left(\alpha-\beta-\gamma+\alpha\beta-\beta\gamma\right)-\eta,\\
\nu &=\frac{1}{2}\left(\alpha +\beta +\gamma +\alpha \gamma +\beta \gamma \right)+\delta +\eta,
\end{align}
The solutions are constructed in the terms of two independent confluent Heun functions as follow,
\begin{equation}
\psi_H=A\operatorname{HeunC}\left(\alpha ,\beta ,\gamma ,\delta ,\eta ,x\right)+Bx^{-\beta}\operatorname{HeunC}\left(\alpha ,-\beta ,\gamma ,\delta ,\eta ,x\right). \label{canonincalheun}
\end{equation}
There is a known polynomial condition for the confluent Heun function to become an $n_r-th$ order polynomial is given as follows,
 \begin{equation}
         \frac{\delta}{\alpha}+\frac{\beta +\gamma}{2}+1=-n_r,\quad n_r\in\mathbb{Z}. \label{HeunPol}
 \end{equation}
One can express confluent Heun's differential equation in its normal form by recognizing $p$ and $q$ functions (following Appendix \ref{AppendixA}),
\begin{align}
p&=\alpha +\frac{\beta +1}{x}+\frac{\gamma +1}{x-1} \ \ , \ \  q=\frac{\mu }{x}+\frac{\nu }{x-1}, \\
\psi_H&=\Psi_H (x)e^{-\frac{1}{2}\alpha x}x^{-\frac{1}{2}\left(\beta+1\right)}{\left(x-1\right)}^{-\frac{1}{2}\left(\gamma +1\right)}.
\end{align}
Now, we evaluate the normal form's $ -\frac{1}{2}\frac{dp}{dx}-\frac{1}{4}p^2+q$ as follows,
\begin{align}
    -\frac{1}{2}\frac{dp}{dx} &= \frac{1}{x^2}\left(\frac{\beta +1}{2}\right)+\frac{1}{{\left(x-1\right)}^2}\left(\frac{\gamma +1}{2}\right), \\
-\frac{1}{4}p^2 &= -\frac{\alpha^2}{4}-\frac{1}{x^2}\left(\frac{\beta^2+1+2\beta}{4}\right)-\frac{1}{{\left(x-1\right)}^2}\left(\frac{\gamma^2+1+2\gamma}{4}\right) \nonumber \\
&~~~~-\frac{2}{x}\left(\frac{\alpha \beta +\alpha}{4}\right)-\frac{2}{x-1}\left(\frac{\alpha \gamma +\alpha}{4}\right)-\frac{2}{x\left(x-1\right)}\left(\frac{\beta \gamma +1+\beta +\gamma}{4}\right).
\end{align}
Combining all terms, we obtain the confluent Heun equation's normal form as follows,
\begin{equation}
    \frac{d^2\Psi_H}{dx^2}+\left(-\frac{\alpha^2}{4}+\frac{\frac{1}{2}-\eta}{x}+\frac{\frac{1}{4}-\frac{\beta^2}{4}}{x^2}+\frac{-\frac{1}{2}+\delta +\eta}{x-1}+\frac{\frac{1}{4}-\frac{\gamma^2}{4}}{{\left(x-1\right)}^2}\right)\Psi_H=0,
\end{equation}
The general solution to this equation can be expressed in term of confluent Heun function. This reads
\begin{multline}
    \Psi_H=e^{\frac{1}{2}\alpha x}x^{\frac{1}{2}\left(\beta +1\right)}{\left(x-1\right)}^{\frac{1}{2}\left(\gamma +1\right)}\times\\\left[A\operatorname{HeunC}\left(\alpha ,\beta ,\gamma ,\delta ,\eta ,x\right)+Bx^{-\beta}\operatorname{HeunC}\left(\alpha ,-\beta ,\gamma ,\delta ,\eta ,x\right)\right].
\end{multline}
In the asymptotically far region, $x\to\infty$, the approximate solution of the canonical Heun differential equation is given as follows,
\begin{align}
    \psi_{H\infty}&=A_\infty x^{-\left(\frac{\delta}{\alpha}+\frac{\beta+\gamma+2}{2}\right)}+B_\infty e^{-\alpha x}x^{\left(\frac{\delta}{\alpha}-\frac{\beta+\gamma+2}{2}\right)}\nonumber\\
    &=e^{-\frac{1}{2}\alpha x}x^{-\frac{\beta+\gamma+2}{2}}\left[A_\infty e^{\frac{1}{2}\alpha x}x^{-\frac{\delta}{\alpha}}+B_\infty e^{-\frac{1}{2}\alpha x}x^{\frac{\delta}{\alpha}}\right],
\end{align}
or, in the normal form, we get,
\begin{equation}
    \Psi_{H\infty}=A_\infty e^{\frac{1}{2}\alpha x}x^{-\frac{\delta}{\alpha}}+B_\infty e^{-\frac{1}{2}\alpha x}x^{\frac{\delta}{\alpha}}. \label{finalsolinfty}
\end{equation}

\section{The Confluent Hypergeometric Equation}\label{AppendixC}
The confluent hypergeometric differential equation is a linear second order ordinary differential equation having this following form \cite{Bell},  
\begin{gather}
    \frac{d^2\psi_C}{dx^2}+\left(-\frac{1}{4}+\frac{k}{x}+\frac{\frac{1}{4}-m^2}{x^2}\right)\psi_C=0. \label{normalformhypergeometric}
\end{gather}
The solutions can be expressed in the terms of two independent Whittaker confluent hypergeometric functions, $M_{k,m}(x)$, as follow,
\begin{equation}
\psi_C=A M_{k,m}(x) +B M_{k,-m}(x).
\end{equation}
The Whittaker confluent hypergeometric functions are connected with the confluent hypergeometric functions, $F_{1,1}\left(a,b,x\right)$, as follows,
\begin{gather}
 M_{k,\pm m}=x^{\frac{1}{2}\pm m}e^{-\frac{x}{2}}F_{1,1}\left(\frac{1}{2}-k\pm m,1\pm 2m,x\right).
\end{gather}
In addition, the confluent hypergeometric function in an asymptotic region i.e., $x\to\infty$ can be written as
\begin{multline}
F_{1,1}\left(a,b,|x|\to\infty\right)=\frac{\Gamma(b)}{\Gamma(a)}e^xx^{a-b}F_{2,0}\left(b-a,1-a,-\frac{1}{x}\right)\\+  \frac{\Gamma(b)}{\Gamma(b-a)}(-x)^{-a}F_{2,0}\left(a,a-b+1,-\frac{1}{x}\right), \label{faratfar}
\end{multline}
where $F_{2,0}\left(a,b,x\right)$ is the hypergeometric function type $\{2,0\}$ \cite{NIST}.

\section{The Gauss Hypergeometric Equation}\label{AppendixD}
The Gauss hypergeometric differential equation is a linear second order ordinary differential equation having this following canonical form \cite{Bell},  
\begin{gather}
    x(1-x)\frac{d^2\psi_G}{dx^2}+\left(a_3-\left(a_1+a_2+1\right)x\right)\frac{d\psi_G}{dx}-a_1a_2\psi_G=0. 
\end{gather}
The solutions of the above equation are given in the terms of the Gauss hypergeometric functions, $F_{2,1}(a_1,a_2,a_3,x)$, as follow,
\begin{equation}
\psi_C=A F_{2,1}(a_1,a_2,a_3,x) +B x^{1-a_3} F_{2,1}(a_1-a_3+1,a_2-a_3+1,2-a_3,x).
\end{equation}
The normal form of the Gauss hypergeometric differential equation and its solutions are given as the following,
\begin{gather}
 \frac{d^2\Psi_G}{dx^2}-\left[\frac{x^2 \left\{(a_1-a_2)^2-1\right\}-2 x \left\{(a_1+a_2-1)a_3-2a_1 a_2\right\} a_3+a_3(a_3-2) }{4 (x-1)^2 x^2}\right]\Psi_G=0, \label{normalformgauss}\\
\Psi_G= x^{\frac{a_3}{2}}(1-x)^{\frac{1}{2}\left(1+a_1+a_2-a_3\right)} \psi_G.
\end{gather}
The following linear second order differential equation also has solutions in the terms of the Gauss hypergeometric functions,
\begin{gather}
    x(1+x)\frac{d^2\psi_G}{dx^2}+\left(2x+1\right)\frac{d\psi_G}{dx}+\left\{\frac{\mathcal{A}}{x}-\frac{\mathcal{B}}{x+1}+\mathcal{C}\right\}\psi_G=0,\label{gauss1}
\end{gather}
with the general solution,
\begin{multline}
 \psi_G= (x+1)^{-i\sqrt{\mathcal{B}}}\left[A x^{i\sqrt{\mathcal{A}}} F_{2,1}\left( \mathcal{A}_1,\mathcal{A}_2,\mathcal{A}_3,-x     \right)\right.\\ \left.+B x^{-i\sqrt{\mathcal{A}}} F_{2,1}\left( \mathcal{A}_4,\mathcal{A}_5,\mathcal{A}_6,-x     \right)\right]   ,
\end{multline}
where,
\begin{align}
   \mathcal{A}_1&=\frac{1}{2}\left(1+\sqrt{1-4\mathcal{C}}\right)+i\left(\sqrt{\mathcal{A}}-\sqrt{\mathcal{B}}\right),\\
   \mathcal{A}_2&=\frac{1}{2}\left(1-\sqrt{1-4\mathcal{C}}\right)+i\left(\sqrt{\mathcal{A}}-\sqrt{\mathcal{B}}\right),\\
   \mathcal{A}_3&=1+2i\sqrt{\mathcal{A}},\\
    \mathcal{A}_4&=\frac{1}{2}\left(1+\sqrt{1-4\mathcal{C}}\right)-i\left(\sqrt{\mathcal{A}}+\sqrt{\mathcal{B}}\right),\\
    \mathcal{A}_5&=\frac{1}{2}\left(1-\sqrt{1-4\mathcal{C}}\right)-i\left(\sqrt{\mathcal{A}}+\sqrt{\mathcal{B}}\right),\\
   \mathcal{A}_6&=1-2i\sqrt{\mathcal{A}}.
\end{align}
The Gauss hypergeometric function has a very useful property that allow us to probe the solution in an asymptotic limit,
\begin{align}
F_{2,1}(a_1,a_2,a_3,x) &= \frac{\Gamma(a_2-a_1)\Gamma(a_3)}{\Gamma(a_2)\Gamma(a_3-a_1)} (-x)^{-a_1}F_{2,1}\left(a_1,a_1-a_3+1,a_1-a_2+1,\frac{1}{x}\right) \nonumber  \\
&~~~+\frac{\Gamma(a_1-a_2)\Gamma(a_3)}{\Gamma(a_1)\Gamma(a_3-a_2)}(-x)^{-a_2} F_{2,1}\left(a_2,a_2-a_3+1,a_2-a_1+1,\frac{1}{x}\right) . \label{connection}
\end{align}
\end{appendices}

\bibliography{sn-bibliography}

\end{document}